\documentstyle[twoside,psbox]{article}

\catcode`\@=11
\long\def\@makefntext#1{
\protect\noindent \hbox to 3.2pt {\hskip-.9pt  
$^{{\eightrm\@thefnmark}}$\hfil}#1\hfill}               

\def\@makefnmark{\hbox to 0pt{$^{\@thefnmark}$\hss}}    
        
\def\ps@myheadings{\let\@mkboth\@gobbletwo
\def\@oddhead{\hbox{}
\rightmark\hfil\eightrm\thepage}   
\def\@oddfoot{}\def\@evenhead{\eightrm\thepage\hfil
\leftmark\hbox{}}\def\@evenfoot{}
\def\sectionmark##1{}\def\subsectionmark##1{}}

\oddsidemargin=\evensidemargin
\newcounter{sectionc}\newcounter{subsectionc}\newcounter{subsubsectionc}
\renewcommand{\section}[1] {\vspace{12pt}\addtocounter{sectionc}{1} 
\setcounter{subsectionc}{0}\setcounter{subsubsectionc}{0}\noindent 
        {\tenbf\thesectionc. #1}\par\vspace{5pt}}
\renewcommand{\subsection}[1] {\vspace{12pt}\addtocounter{subsectionc}{1} 
        \setcounter{subsubsectionc}{0}\noindent 
        {\bf\thesectionc.\thesubsectionc. {\kern1pt \bfit #1}}\par\vspace{5pt}}
\renewcommand{\subsubsection}[1] {\vspace{12pt}\addtocounter{subsubsectionc}{1}
        \noindent{\tenrm\thesectionc.\thesubsectionc.\thesubsubsectionc.
        {\kern1pt \tenit #1}}\par\vspace{5pt}}
\newcommand{\nonumsection}[1] {\vspace{12pt}\noindent{\tenbf #1}
        \par\vspace{5pt}}

\newcounter{appendixc}
\newcounter{subappendixc}[appendixc]
\newcounter{subsubappendixc}[subappendixc]
\renewcommand{\thesubappendixc}{\Alph{appendixc}.\arabic{subappendixc}}
\renewcommand{\thesubsubappendixc}
        {\Alph{appendixc}.\arabic{subappendixc}.\arabic{subsubappendixc}}

\renewcommand{\appendix}[1] {\vspace{12pt}
        \refstepcounter{appendixc}
        \setcounter{figure}{0}
        \setcounter{table}{0}
        \setcounter{lemma}{0}
        \setcounter{theorem}{0}
        \setcounter{corollary}{0}
        \setcounter{definition}{0}
        \setcounter{equation}{0}
        \renewcommand{\thefigure}{\Alph{appendixc}.\arabic{figure}}
        \renewcommand{\thetable}{\Alph{appendixc}.\arabic{table}}
        \renewcommand{\theappendixc}{\Alph{appendixc}}
        \renewcommand{\thelemma}{\Alph{appendixc}.\arabic{lemma}}
        \renewcommand{\thetheorem}{\Alph{appendixc}.\arabic{theorem}}
        \renewcommand{\thedefinition}{\Alph{appendixc}.\arabic{definition}}
        \renewcommand{\thecorollary}{\Alph{appendixc}.\arabic{corollary}}
        \renewcommand{\theequation}{\Alph{appendixc}.\arabic{equation}}
        \noindent{\tenbf Appendix \theappendixc #1}\par\vspace{5pt}}
\newcommand{\subappendix}[1] {\vspace{12pt}
        \refstepcounter{subappendixc}
        \noindent{\bf Appendix \thesubappendixc. {\kern1pt \bfit #1}}
        \par\vspace{5pt}}
\newcommand{\subsubappendix}[1] {\vspace{12pt}
        \refstepcounter{subsubappendixc}
        \noindent{\rm Appendix \thesubsubappendixc. {\kern1pt \tenit #1}}
        \par\vspace{5pt}}

\newcommand{\textlineskip}{\baselineskip=13pt}
\newcommand{\smalllineskip}{\baselineskip=10pt}

\def\eightcirc{
\begin{picture}(0,0)
\put(4.4,1.8){\circle{6.5}}
\end{picture}}
\def\eightcopyright{\eightcirc\kern2.7pt\hbox{\eightrm c}} 

\newcommand{\copyrightheading}[1]
        {\vspace*{-2.5cm}\smalllineskip{\flushleft
        {\footnotesize International Journal of Modern Physics A, #1}\\
        {\footnotesize $\eightcopyright$\, World Scientific Publishing
         Company}\\
         }}

\newcommand{\publisher}[2]{{\begin{center}\footnotesize\smalllineskip 
        Received #1\\
        Revised #2
        \end{center}
        }}

\def\abstracts#1#2#3{{
        \centering{\begin{minipage}{4.5in}\baselineskip=10pt\footnotesize
        \parindent=0pt #1\par 
        \parindent=15pt #2\par
        \parindent=15pt #3
        \end{minipage}}\par}} 

\newcommand{\bibit}{\nineit}
\newcommand{\bibbf}{\ninebf}
\renewenvironment{thebibliography}[1]
        {\frenchspacing
         \ninerm\baselineskip=11pt
         \begin{list}{\arabic{enumi}.}
        {\usecounter{enumi}\setlength{\parsep}{0pt}
         \setlength{\leftmargin 12.7pt}{\rightmargin 0pt} 
         \setlength{\itemsep}{0pt} \settowidth
        {\labelwidth}{#1.}\sloppy}}{\end{list}}

\newcounter{itemlistc}
\newcounter{romanlistc}
\newcounter{alphlistc}
\newcounter{arabiclistc}

\newcommand{\fcaption}[1]{
        \refstepcounter{figure}
        \setbox\@tempboxa = \hbox{\footnotesize Fig.~\thefigure. #1}
        \ifdim \wd\@tempboxa > 5in
           {\begin{center}
        \parbox{5in}{\footnotesize\smalllineskip Fig.~\thefigure. #1}
            \end{center}}
        \else
             {\begin{center}
             {\footnotesize Fig.~\thefigure. #1}
              \end{center}}
        \fi}

\newcommand{\tcaption}[1]{
        \refstepcounter{table}
        \setbox\@tempboxa = \hbox{\footnotesize Table~\thetable. #1}
        \ifdim \wd\@tempboxa > 5in
           {\begin{center}
        \parbox{5in}{\footnotesize\smalllineskip Table~\thetable. #1}
            \end{center}}
        \else
             {\begin{center}
             {\footnotesize Table~\thetable. #1}
              \end{center}}
        \fi}

\def\@citex[#1]#2{\if@filesw\immediate\write\@auxout
        {\string\citation{#2}}\fi
\def\@citea{}\@cite{\@for\@citeb:=#2\do
        {\@citea\def\@citea{,}\@ifundefined
        {b@\@citeb}{{\bf ?}\@warning
        {Citation `\@citeb' on page \thepage \space undefined}}
        {\csname b@\@citeb\endcsname}}}{#1}}

\newif\if@cghi
\def\cite{\@cghitrue\@ifnextchar [{\@tempswatrue
        \@citex}{\@tempswafalse\@citex[]}}
\def\citelow{\@cghifalse\@ifnextchar [{\@tempswatrue
        \@citex}{\@tempswafalse\@citex[]}}
\def\@cite#1#2{{$\null^{#1}$\if@tempswa\typeout
        {IJCGA warning: optional citation argument 
        ignored: `#2'} \fi}}

\def\pmb#1{\setbox0=\hbox{#1}
        \kern-.025em\copy0\kern-\wd0
        \kern.05em\copy0\kern-\wd0
        \kern-.025em\raise.0433em\box0}

\def\fnt#1#2{\footnotetext{\kern-.3em
        {$^{\mbox{\scriptsize #1}}$}{#2}}}

\def\fpage#1{\begingroup
\voffset=.3in
\thispagestyle{empty}\begin{table}[b]\centerline{\footnotesize #1}
        \end{table}\endgroup}

\def\runninghead#1#2{\pagestyle{myheadings}
\markboth{{\protect\footnotesize\it{\quad #1}}\hfill}
{\hfill{\protect\footnotesize\it{#2\quad}}}}
\font\tenrm=cmr10
\font\tenit=cmti10 
\font\tenbf=cmbx10
\font\bfit=cmbxti10 at 10pt
\font\ninerm=cmr9
\font\nineit=cmti9
\font\ninebf=cmbx9
\font\eightrm=cmr8

\def\qed{\hbox{${\vcenter{\vbox{                        
   \hrule height 0.4pt\hbox{\vrule width 0.4pt height 6pt
   \kern5pt\vrule width 0.4pt}\hrule height 0.4pt}}}$}}

\newif\ifpreprintsty \global\preprintstyfalse

\def\@chuckoptarg[#1]{}

\newif\if@floats \@floatsfalse
\def\ds@floats{\@floatstrue}

\def\fnum@figure{\figurename\penalty10000\hskip.3em plus .1em\relax\thefigure.}
\if@floats
\def\figure{\let\@capwidth\columnwidth\@float{figure}}
\let\endfigure\end@float
\@namedef{figure*}{\let\@capwidth\textwidth\@dblfloat{figure}}
\@namedef{endfigure*}{\end@dblfloat}
\else
\def\figure{%
\let\@capwidth\columnwidth
\ifpreprintsty\iffirstfig
{\newpage\centerline{FIGURES}}\global\firstfigfalse
\fi
\vskip1pc
\def\@captype{figure}%
\interlinepenalty10000 %
\@ifnextchar[{\@chuckoptarg}{}%
}%
\def\endfigure{\goodbreak\vskip1pc}%
\@namedef{figure*}{\figure}%
\@namedef{endfigure*}{\endfigure}%
\fi

\begin{document}

\runninghead{DNA Transcription Mechanism}
{Julian J.-L. Ting}

\normalsize\textlineskip
\thispagestyle{empty}
\setcounter{page}{1}

\copyrightheading                    {Vol. 7, No. 5 (1997) 1125--1132}

\vspace*{0.88truein}

\fpage{1}
\centerline{\bf DNA Transcription Mechanism with a Moving Enzyme}
\vspace*{0.37truein}
\centerline{\footnotesize Julian Juhi-Lian Ting}
\vspace*{0.015truein}
\centerline{\footnotesize\it
Institute of Atomic and Molecular Sciences, 
Academia Sinica, P.O.Box 23-166, Taipei, Taiwan 106, R.O.C.
\footnote{E-mail address:jlting@gate.sinica.edu.tw}
}
\vspace*{0.225truein}
\publisher{(\today)}{(revised date)}

\vspace*{0.21truein}
\abstracts{
Previous numerical investigations of an
one-dimensional DNA model with an extended modified coupling constant by
transcripting enzyme are integrated to longer time and demonstrated explicitly
the trapping of breathers by DNA chains with realistic parameters obtained from experiments.
Furthermore, collective coordinate method is used to 
explain a previously observed
numerical evidence that breathers placed far from defects are difficult to
trap, and the motional effect of RNA-polymerase is investigated.
}{}{}

\textlineskip                  
\vspace*{12pt}                 

\section{Introduction}

The editors of {\it Science}, 
Culotta \& Koshland [1994],
announced in the
last issue of 1994 that `the molecule of 1994' is DNA.
One can still find many papers in the physical literature about DNA recently.
However, most physicists are concerned about its statistical properties
by analyzing some database of DNA sequences, for instance
Azbel' [1995] found DNA sequences have no long-range correlations,
instead of dynamical properties analyzed below.

\begin{figure}[hbt]
\begin{center}
\mbox{\psboxto(7cm;6cm){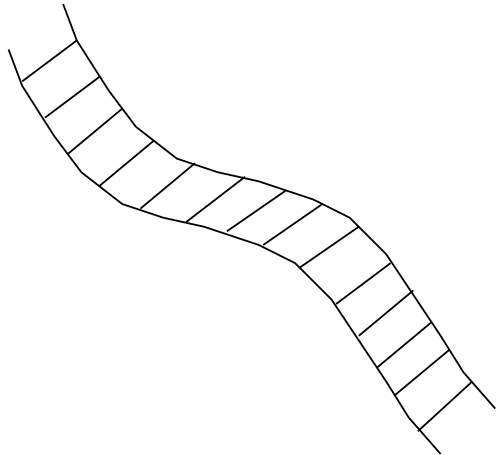}} \\
\caption[1]{An artistic plot of DNA.}
\label{DNA}
\end{center}
\end{figure}

An artistic plot of  DNA which appeared
on the cover page of that  issue of {\it Science} and most textbooks of 
biochemistry is 
similar to Fig.~\ref{DNA}.
As one can see, it is a double-helix with hydrogen bonds
connecting those two strands.
Of course, this is a simplified picture.
However, Peyrard \& Bishop [1989] untwisted
this ladder like DNA,
and wrote down its equation of motion.

Biologists told us that transcription  of RNA is the first step 
in a chain of events leading to expression of the genetic information
encoded in double-stranded DNA. During this process 
RNA-polymerase will attached to DNA.
Erie {\it et al.} [1993] consider
the role of RNA-polymerase in transcription
is to synthesize, under the DNA template, 
the nascent RNA chain with high fidelity and at
reasonable rates.
However, for physicists the enzyme may play an additional role {\it before}
transcription, i.e. to focus thermal energy so that the double-helix 
can be opened.

Englander {\it et al.} [1980] have proposed the existence of solitons on DNA
helices more than 15 years ago.
Technical difficulties prevent direct observation of solitons
on DNA.
However, 
Urabe \& Tominaga [1981] have
Raman spectra showing mode softening which might
be accounted by the existence of breather traveling.
On the other hand, since DNA is a kind of polymer and solitons have been 
recognized as important conformational excitations in polymers. 
Furthermore, numerical and analytical investigations of the Peyrard-Bishop
model of DNA, which can be rationally deduced from a real DNA 
atomic structure, show the existence of breathers traveling on  
DNA chains. We are now more confident about the existence of breathers
on DNA chains today.

Forinash {\it et al} [1991,1994] considered
the interaction of enzyme with DNA during transcription as an isolated
impurity.
In a previous paper, Ting \& Peyrard [1996] considered
numerically the possibility of breather
trapping by a continuous defect and by
a multiple scale analysis the necessary condition for
trapping to be a reduced coupling constant within the defect area.
The conclusion was loose about whether the breather will be
trapped by a real DNA. Due to limited computing resources, numerical
integrations were only carried out to about 5000 time steps.
However, Dauxois \& Peyrard [1993] indicated that DNA is a discrete object
and the behaviour of a breather at this region of coupling is mainly
governed by discrete effect.
In that case a stable breather should be thin and move slowly.
It is the purpose of this paper
to extend those numerical integration to longer time to investigate
the long time behaviour of that kind of breather motion.
The second purpose of this paper is to extend previous collective coordinate
method to show the difficulty of trapping of a far away breather and
to consider the effect of moving enzyme, because Erie {\it et al.} [1993] told
us that RNA-polymerase is actually moving.

\section{The Peyrard-Bishop DNA Model}

\begin{figure}[bth]
\begin{center}
\mbox{\psboxto(7cm;7cm){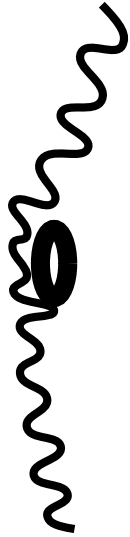}} \\
\caption[1]{An artistic plot of a DNA with enzyme, showing the enzyme
contact DNA at multiple sites and bend DNA towards itself. }
\label{bend}
\end{center}
\end{figure}

The Peyrard-Bishop model 
considers a harmonic coupling for neighboring 
sites and within
the same strand, while an anharmonic Morse coupling
between strands.
A transformation into normal coordinate is made and 
the longitudinal mode thrown away, because it is too small.
The equations obtained,  in dimensionless form, read
\begin{eqnarray}
\label{ND}
  {{\partial^2 y_n} \over {\partial t^2}} - 
  k_{n+1} ( y_{n+1} - y_n ) + 
  k_n ( y_n - y_{n-1} ) 
  - 2  e^{-  y_n} 
  ( e^{-  y_n} - 1)  = 0 \; ,
\end{eqnarray}
in which $y_n$ is the relative displacement between strands and $k_n$
is the coupling constant between $y_n$ and $y_{n-1}$.
Because we have untwisted the double-helix, there is no difference 
between A-, B- or Z- types of DNA. Furthermore, the present model does not
taken specific DNA sequence into account.
Ansari {\it et al.} [1995]
told us, protein contact DNA at certain  
sites and bend DNA towards itself as
shown in Fig.~\ref{bend}.
This will results in a reduced coupling constant for the outside strand.
Presumably, enzyme can function as  a kind of lens to focus some
thermo fluctuation waves present in normal physiological condition,
traveling along the strand.

\subsection{Long-time numerical integrations}

Previous numerical integrations can be summarized as the
following:
Taller breathers, hence thinner, can be trapped by a defect with
reduced coupling constant,
while shorter breathers are broader and result in only concentrated
energy and pass through.
Therefore trapping can be controlled by breather's amplitude.
Furthermore, these two points indicate breathers move collectively as rigid
bodies just like bike wheels.

\begin{figure}[bth]
\begin{center}
\mbox{\psboxto(15cm;15.5cm){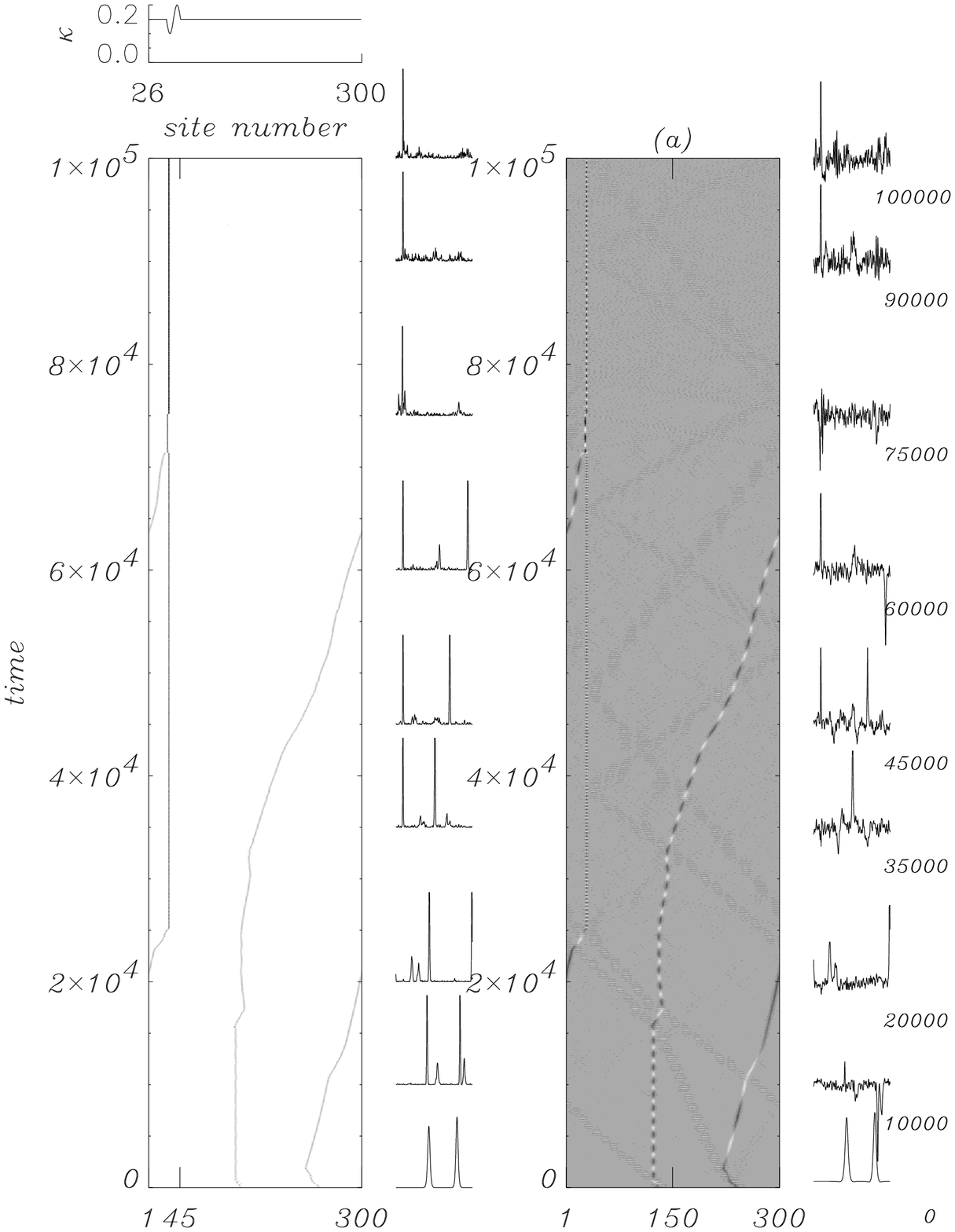}} \\
\mbox{\psboxto(15cm;15.5cm){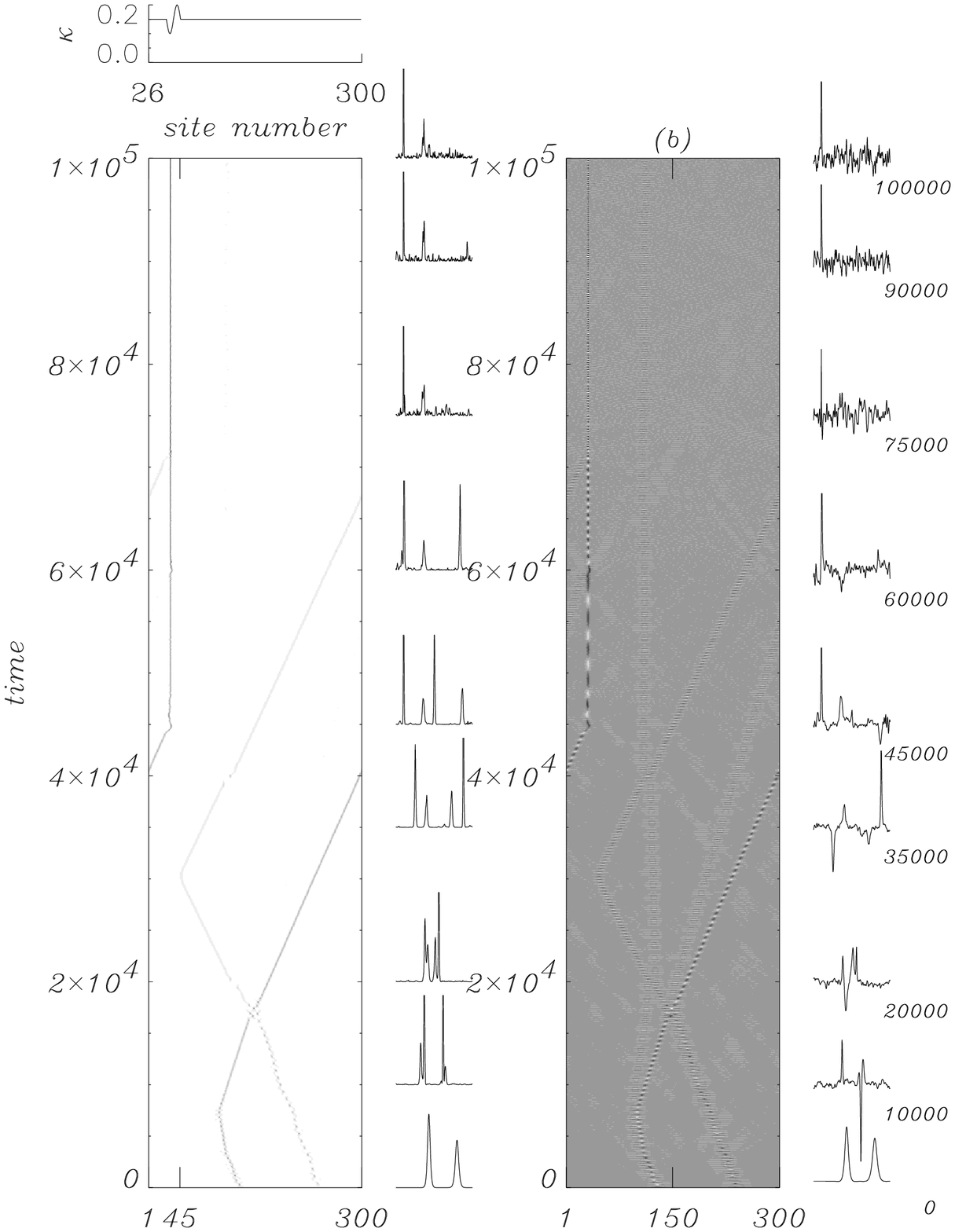}} \\
\caption{Half-tone plots of the energy distribution of 
breather evolutions, left, and the corresponding amplitudes, right, from 
direct numerical integration of Eq.~(\protect\ref{ND}).
In each figure the top insert
shows the variation of the coupling constant used for the calculation, while
the right inserts show snap-shots of the breather 
energy or amplitude distributions.
Defects positions are shown in the plots axis. 
The breathers start from $X ( 0 ) = 130, 240$, 
$K_n = 0.15$ outside the defect, the amplitude of sinusoidal 
defect is $0.05$ and (a) $u_c = 0.14, 0.13$, $u_e = -0.05, -0.06$.
(b) $u_c = 0.13, 0.11$, $u_e = -0.04, -0.03.$
}
\label{real}
\end{center}
\end{figure}

\begin{figure}[bth]
\begin{center}
\mbox{\psboxto(15cm;15.5cm){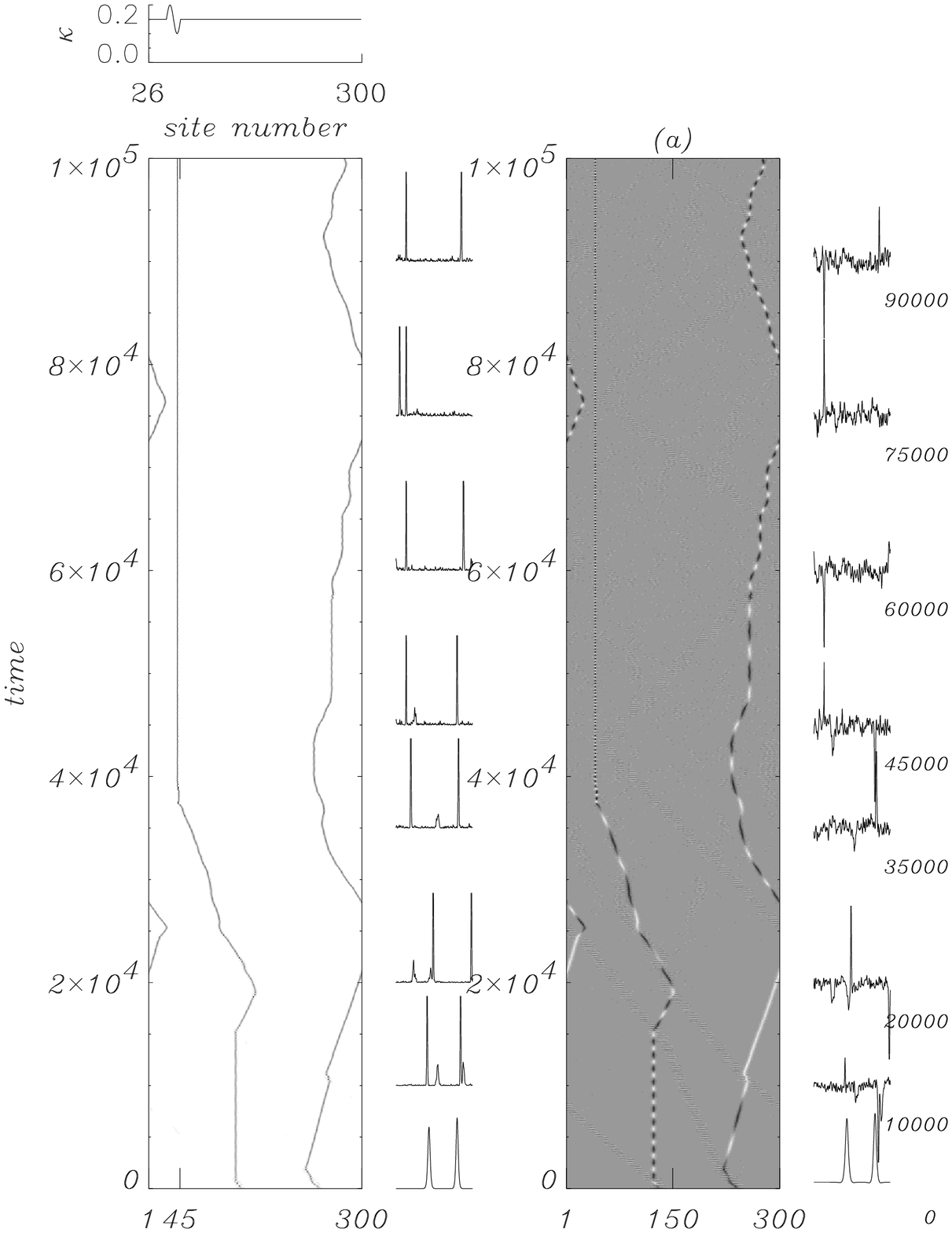}} \\
\mbox{\psboxto(15cm;15.5cm){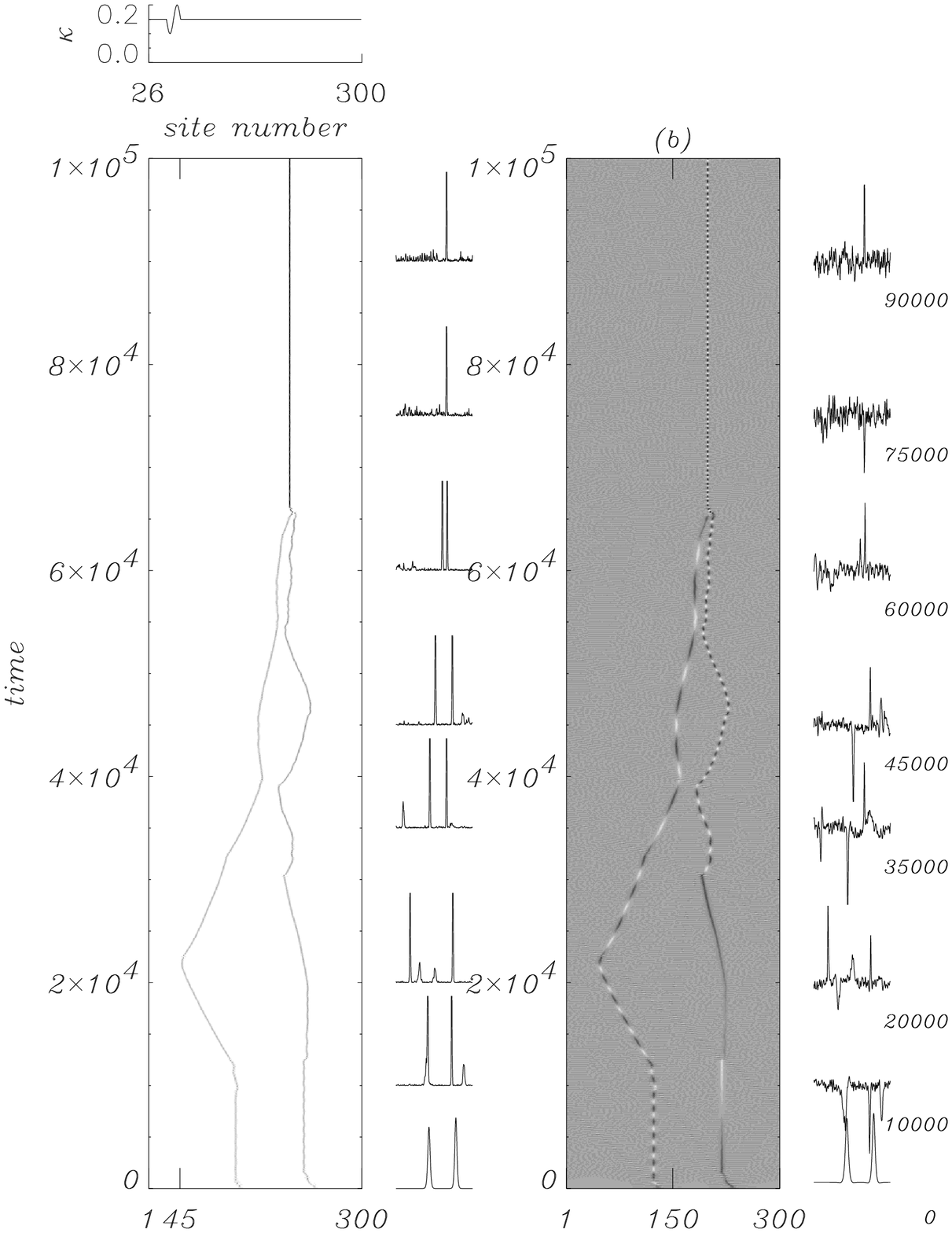}} \\
\caption{Both figures are the same as Fig.~\protect\ref{real}(a), except
(a) the defect pattern is reversed.
(b) the initial breather position are closer. 
}
\label{easy}
\end{center}
\end{figure}

One may wonder whether the results obtained in the previous paper 
are applicable to more realistic cases.  
In real DNA one has $k_n \approx 0.13$ to
$0.15$. 
Two pictures for trapping by realistic DNA are shown in Fig~\ref{real}.
These breathers are 
so weak, i.e. move so slowly, so that they are vulnerable to the presence of 
defect or
other breathers. In Fig.~\ref{easy}(a) the defect pattern is reversed 
in comparison with Fig.~\ref{real}(a). But we found quite different picture for
the breather motion and they were attracted by the region of 
lowered coupling constant. Fig.~\ref{easy}(b) shows two
breathers, the same as in Fig.~\ref{real}(a) except for their separation,
attracting  each other.

\subsection{Collective coordinate method for moving defect}

Following Remoissenet [1986],
a slightly different derivation from the one used before is
used below to consider the effect of moving enzyme.
Firstly, in the continuum limit with a Taylor expansion in the 
potential term which assumes small amplitude excitation Eq.~(\ref{ND}) became,

\begin{equation}
\label{continue}
   {{\partial^2 y} \over {\partial t^2}}
  - {\partial \over {\partial x} } (k_1
  {{\partial y} \over {\partial x}}  ) d^2
  + 2   ( y -
  {3 \over 2} y^2 +
  {7 \over 6} y^3 ) = 0 \; .
\end{equation}
Because of the small oscillation assumption, we have
$y \approx \epsilon \phi$ and
\begin{equation}
\phi = F_1 e^{i \theta} + F_1^* e^{- i \theta} + 
\epsilon ( F_0 + F_2 e^{ 2 i \theta} + F_2^* e^{ - 2 i \theta} ) + O (
\epsilon^2 ) \; ,
\end{equation}
in which
$F_1 = F_1 (x_1, t_1)$ and $F_2 = F_2 ( x_1, t_1)$ with $x_1 =
\epsilon x$ etc..
This expansion is equivalent to 
\begin{eqnarray}
\phi & = & F_0 + \epsilon F_1 + \epsilon^2 F_2 + O (\epsilon^3) \; 
\end{eqnarray}
used before.
Furthermore, for low frequency breathers we considered, we have
$\theta \approx - \omega t$, and
\begin{eqnarray}
{\partial \over {\partial x}} &  = &
{\partial \over {\partial x_0}} + {\partial \over {\partial x_1}}
\epsilon +
O ( \epsilon ^2 ) \; , \\
{\partial^2 \over {\partial x^2}} & = &
{\partial^2 \over {\partial x_0^2}} + 2 { \partial^2 \over 
{\partial x_0 \partial x_1}}
\epsilon +
O ( \epsilon ^2 ) \; .
\end{eqnarray}
We also assumed 
\begin{equation}
\label{assume}
{{\partial k_1}  \over {\partial x}} \approx
{{\partial k_1} \over {\partial x_1}} \; \epsilon \; ,
\end{equation}
for the smooth variation of the defect.
Collecting terms of equal powers in $e^{i \omega t}$, with
$\omega^2 \approx \omega_0^2 = 2$,
we obtain
\begin{eqnarray}
F_0 - 3 F_1 F_1^* = 0 \; , \nonumber \\
3 \epsilon \alpha  \omega^2 ( F_0 F_1 + F_1^* F_2 ) - {7 \over 2}
\epsilon \alpha^2 \omega^2 F_1^* F_1^2 + k \epsilon
{{\partial^2 F_1} \over \partial x_1^2} + 2 i \omega  
{{\partial F_1} \over {\partial t_1}}
- \epsilon {{\partial^2 F_1} \over {\partial t_1^2}}
+ {{\partial k} \over {\partial x}} {{\partial F_1} \over {\partial
x_1}} = 0 \;, \nonumber \\
F_2 =  - {1 \over 2} F_1^2 \; . \nonumber
\end{eqnarray}
From the above three equations we obtained a 
perturbed Nonlinear Schr\"odinger equation (NLS)
at order $\epsilon^2$:
\begin{equation}
 2 i \omega { \partial F_1 \over \partial t_2} +
{{\partial k_1} \over {\partial x_1}}
{{\partial F_1} \over {\partial x_1}} +
k_1 {{\partial^2 F_1} \over {\partial x_1^2}}
+ 8  F_1 | F_1 |^2 = 0 \;.
\end{equation}
We can rescale it into a standard form,
\begin{equation}
  i  {u}_{\hat t}  + {1 \over 2} u_{\hat x \hat x} +
  u | u |^2 + {1 \over 2}
{{\partial } \over {\partial {\hat x}}} ( {\hat k}
{u}_{\hat x} ) = 0 \;,
\end{equation}
and obtain the corresponding Lagrangian density,
\begin{equation}
\Lambda = {i \over 2} ( {u}^* {u}_{\hat t} -
{u} {u}_{\hat t}^*)
- {1 \over 2} ( { 1 }  + {\hat k} ) | {u}_{\hat x} |^2
+ {1 \over 2} | {u} | ^ 4 \; .
\end{equation}
A collective coordinate {\it ansatz},
\begin{equation}
\label{ansatz}
u ( x , t ) = \eta\, {\rm sech} ( {{\eta x} } - \zeta )\,
e^{i (\phi + \xi x )} \;,
\end{equation}
brings the full space integrated Lagrangian density into
\begin{eqnarray}
\label{eqlagrange}
L &=& - 2 \eta \phi_t - 2  \zeta \xi_t +
{{ \eta^3 } \over 3}  -  \xi^2 \eta
 - {1 \over 2} \int_{- \infty}^{+ \infty}  k | u_x |^2 d x
\; .
\end{eqnarray}
For a moving defect
\begin{equation}
\label{step}
 k = \kappa [ \Theta ( x -v t + l ) - \Theta ( x - v t - l ) ] \; ,
\end{equation}
we have the equation of motion for the collective coordinate variables
\begin{eqnarray}
\phi_t &=& {{ {\eta^2} \over 2} }  - { {\xi^2 \over 2} }
 - { \kappa \over 4} ( T_+ + T_- ) \xi^2
- {{\kappa } \over 4} ( S_+^2 ( l - v t ) + S_-^2 ( l + v t ) ) \xi^2 \eta
\nonumber \\
& &- {{\kappa } \over 4} ( S_+^2T_+^2 ( l - v t ) + S_-^2T_-^2 ( l + v t ) ) \eta^3
- {{ \kappa} \over 4} ( T_+^3 + T_+^3 ) \eta^2 \, ,
\\
\xi_t &=&
- {{\kappa } \over 4} ( S_+^2T_+^2 - S_-^2T_-^2 ) \eta^3
- {{\kappa } \over 4} ( S_+^2 - S_-^2 ) \xi^2 \eta \, , \\
\zeta_t &=&  \xi \eta + {{\kappa} \over 2} ( T_+ + T_- ) \xi \eta \, ,
\label{CZ}\\
\eta_t &=& 0 \label{CA} \, ,
\end{eqnarray}
in which
\begin{eqnarray}
 T_+ &=& -  \tanh ( \eta v t - \eta  l  - \zeta ) \, ,\label{TT}\\
 T_- &=& ~~~  \tanh ( \eta v t + \eta  l  - \zeta ) \, ,\\
 S_+ &=& ~~~  {\rm sech} (\eta v t -  \eta  l  - \zeta ) \, ,\\
 S_- &=& ~~~  {\rm sech} (\eta v t + \eta  l  - \zeta ) \, \label{SM}.
\end{eqnarray}

We find there is nothing essential different from previous stationary defect
located at origin, which can be obtained as a special case of present result
at $t = 0$, except for the variation of $\phi_t$ and the redefinition of 
Eqs.~(\ref{TT})-(\ref{SM}).
We can still prove that a reduced coupling constant in the defect 
area is a necessary condition for a breather to be trapped.
Since a  necessary condition for trapping is $\zeta_t = 0$ more than
twice,  which, according to Eq.~(\ref{CZ}), is equivalent to 
\begin{equation}
\cosh^2 ( \eta v t - \zeta ) =
1 - \cosh^2 ( \eta l ) - {\kappa} \sinh ( \eta l ) \cosh ( \eta l ).
\label{trap}
\end{equation}
Because $\cosh^2 > 1$ and $\eta > 0$, 
$\kappa $ has to be negative.
Furthermore, we can interpret the above equation as:
the greater the separation between the breather and the defect; the
larger the left hand side, the smaller $\kappa$, i.e. the greater 
$ | \kappa |$, should be in order to balance the left hand side to fulfill
the trapping requirement.  This point is consistent with our previous numerical
experience that if we put the initial breathers farther from origin
we require larger breather amplitude for the breather to be trapped.
However, the smallest $\kappa $ is $- 1$.

\section{Conclusion}

We have demonstrated explicitly the possibility for trapping of
realistic DNA. Furthermore, we considered the effect of moving
defect. The result shows no special influence on trapping through
velocity except through changing the breather-defect separation 
distance under our 
collective coordinate model. However,  a
recent measurement by
Yin {\it et al.} [1995] on the RNA-polymerase moving force
might be taken into account by asymmetrical
potentials and stochastic resonance, similar to 
J{\"u}licher \& Prost [1995] for molecular motors.

There are other interesting works on DNA models:
For instance,
Yakushevich [1989] proposed another DNA model.
More models can be found in the references of that paper and 
Forinash's [1991].
Within the Peyrard-Bishop model, Hisakado \& Wadati [1995] have studied
its inhomogeneity effect by considering random mass and random
coupling constant. 
It will also be interesting to take long range columbic interaction
into account as Gaididei {\it et al.} [1995] considered.
Slepyan {\it et al.} [1995] considered solitary waves traveling on a
three dimensional helicoidal fiber, which might be extendible to
a three dimensional DNA model.
However, much work remain to be done in the future.

\section{Acknowledgment}
I thank Professor M. Peyrard for many useful discussions,
the organization committee of  the NLDC international workshop for making the
trip possible,
Drs. Victor W.-K. Wu, Gautam Gangopadhyay and Railing Chang 
for reading the manuscript.

\nonumsection{References}

\end{document}